\documentclass[12pt]{article}

\usepackage{a4wide}
\usepackage{tikzsymbols}
\usepackage{xcolor}
\usepackage{bm}

\DeclareRobustCommand{\orcidicon}{
	\begin{tikzpicture}
	\draw[lime, fill=lime] (0,0) 
	circle [radius=0.2] 
	node[white] {{\fontfamily{qag}\selectfont \tiny ID}};
	\draw[white, fill=white] (-0.0625,0.095) 
	circle [radius=0.007];
	\end{tikzpicture}
	\hspace{-2mm}
}

\foreach \x in {A, ..., Z}{\expandafter\xdef\csname orcid\x\endcsname{\noexpand\href{https://orcid.org/\csname orcidauthor\x\endcsname}
			{\noexpand\orcidicon}}
}

\newcommand{\MeV}{\mathop{\rm MeV}\nolimits}
\newcommand{\sn}{\mathop{\rm sn}\nolimits}
\newcommand{\cn}{\mathop{\rm cn}\nolimits}
\newcommand{\dn}{\mathop{\rm dn}\nolimits}
\newcommand\myeq{\mathrel{\stackrel{\makebox[0pt]{\mbox{\normalfont\tiny reg}}}{=}}}

\begin{document}

\begin{center}
{\Large\bf Non-perturbative quantum Yang--Mills at finite temperature beyond
  lattice: \it{a Dyson--Schwinger approach}}\\[1cm]
{\Large Marco Frasca$^1$, Anish Ghoshal$^2$ and Stefan Groote$^3$}\\[12pt]
$^1$ Rome, Italy\\[7pt]
$^2$ Institute of Theoretical Physics, Faculty of Physics,\\
University of Warsaw, ul.\ Pasteura 5, 02-093 Warsaw, Poland,\\[7pt]
$^3$ Institute of Physics, University of Tartu,
W.~Ostwaldi 1, EE-50411 Tartu, Estonia\\[7pt]
\end{center}
\vspace{1cm}

\begin{abstract}\noindent
Using a Dyson--Schwinger approach, we perform an analysis of the
non-trivial ground state of thermal $SU(N)$ Yang--Mills theory in the
non-perturbative regime where chiral symmetry is dynamically broken by a
mass gap. Basic thermodynamic observables such as energy density and
pressure are derived analytically, using Jacobi elliptic functions. The
results are compared with lattice results. Good agreement is found at low
temperatures, providing a viable scenario of a gas of massive glue states
populating higher levels of the spectrum of the theory. At high temperatures
a scenario without glue states consistent with a massive scalar field is
observed, showing an interesting agreement with lattice data. The
possibility is discussed that the results derived in this analysis open up a
novel pathway beyond lattice to precision studies of phase transitions with
false vacuum and cosmological relics that depend on the equations of state in
strong coupled gauge theories of the type of Quantum Chromodynamics (QCD).
\end{abstract}

\section{Introduction\label{introduction}}
Nowadays, only lattice simulations and other very limited theoretical
approaches can reliably analyze the key observables in quantum Yang--Mills
theories at finite temperatures in the strong coupling regime. Existing
analytic methods such as the Polyakov loop model~\cite{Pisarski:2000eq,%
Sannino:2002wb} still rely on sets of unknown non-perturbative parameters in
the effective potential obtained by properly fitting lattice data. The exact
form of the coefficients of a thermodynamic potential is found by an
appropriate fit to the corresponding lattice data~\cite{Panero:2009tv}, that
fully determines this potential for a given theory. Such a lattice-inspired
effective model has been extensively employed in view of cosmological studies
involving dark sectors (see e.g.\ Refs.~\cite{Carenza:2022pjd}).

In this work, we develop an approach enabling a fully analytical treatment of
non-perturbative quantum field thermodynamics in terms of physical parameters
only. Our technique provides, in principle, a systematic way to incorporate
finite temperature corrections to the exact Green function in the framework of
a Dyson--Schwinger approach with a non-trivial ground
state~\cite{Frasca:2015yva,Frasca:2016sky,Frasca:2017slg,Chaichian:2018cyv}.
We consider a pure Yang--Mills theory, i.e., a gauge theory that has no other
degrees of freedom than its potentials. It should be noted that Yang--Mills
theories appear in nature only in interaction with other fields like fermions
and scalars, such as in the Standard Model or QCD. Therefore, the idealized
situation without other fields we consider here can only be compared to
lattice calculations~\cite{Koberinski:2019tqk,Brink:2015frg}. We
demonstrate the power of our method by computing the effective action of a
pure Yang--Mills quantum thermal theory under the $SU(N)$ symmetry group in
the non-perturbative regime. The latter is then used to estimate some of the
basic thermodynamic observables such as energy density and pressure. This is
achieved by computing the effective energy--momentum tensor of the Yang--Mills
theory and evaluating thermal averages of its components in terms of the
correlation functions~\cite{daSilva:2002dje,Santos:2019xlx}. Comparing the
thermodynamic observables with the lattice data available for $N=3$ or $4$
colors~\cite{Panero:2009tv} reveals an overall consistency of our results,
suggesting that our method captures the most essential features of
non-perturbative Yang--Mills dynamics at low temperatures. A recent approach
with a free gas of glueballs shows a similar consistency~\cite{Trotti:2022knd}.
As pure Yang--Mills theory is not seen in experiments, we emphasize that a
proof of principle of the symmetry breaking in Yang--Mills theory can only
been obtained by comparing with lattice calculations~\cite{Panero:2009tv}. Our
aim is to show how, from calculations based on the full theory, very good fits
can be obtained at low and at high temperatures, providing a sound physical
interpretation for the lattice results.

To properly frame this work, we study a Yang--Mills theory without
interactions with other fields like quarks or scalars. This is done  by
starting with the full theory and developing an exact solution that is
afterwards fitted to the lattice results presented in
Ref.~\cite{Panero:2009tv}. This procedure should prove the soundness of our
technique also for thermal field theory. It should be emphasized that, at low
temperatures, the results presented in Ref.~\cite{Panero:2009tv} are too few
to obtain a significant result beyond $SU(3)$ and $SU(4)$ Yang--Mills theories.
Therefore, we limit our analysis to these cases.

The paper is structured as follows. In Sec.~2, we present our technique to
obtain an exact solution of the quantum Yang--Mills theory. In Sec.~3, we show
how the partition function is obtained in our case. In Sec.~4, we derive all
the thermodynamic variables for the Yang--Mills theory. In Sec.~5, we evaluate
the partition function and present our results in comparison wth lattice data.
Finally, in Sec.~6, we give our conclusions.

\section{Gaussian solution of quantum Yang--Mills theory}
We study an exact solution of the quantum Yang--Mills field theory that
provides a Gaussian partition function~\cite{Frasca:2015yva,Frasca:2016sky,%
Frasca:2017slg,Chaichian:2018cyv,Frasca:2023uaw}. It is worthwhile to study
such a solution because exact solutions in quantum field theory are quite rare
and mostly given for non-physical models only. We expect that the solution
found here represents the behavior of the Yang--Mills theory. In order to
simplify the presentation, we consider a Yang--Mills theory with a $SU(N)$
gauge group. The action at the classical level is given by
\begin{equation}
S_{YM}=\int d^4x\left(-\frac14F_{\mu\nu}^aF^{\mu\nu}_a+j^\mu_aA_\mu^a\right)
\end{equation}
(latin letters $a,b,c,\ldots$ represent group indices taking the values
$1,2,3$), where $j^\mu_a$ is a generic source and an element of the $su(2)$
algebra. The field strength tensor is given by 
\begin{equation}
F_{\mu\nu}^a=\partial_\mu A_\nu^a-\partial_\nu A_\mu^a+gf^{abc}A_\mu^b A_\nu^c,
\end{equation}
where $g$ is the coupling constant and $f^{abc}$ are the structure constants
of the gauge group, leading to the classical equations of motion given by
\begin{equation}\label{eq:YM}
\partial^\mu F_{\mu\nu}^a+gf^{abc}A^{b\mu}F^c_{\mu\nu}=j_\nu^a.
\end{equation}
In the following we will provide a solution to Eq.~(\ref{eq:YM}). The
classical equation of motion generalizes to the quantum case by introducing a
gauge fixing term and ghost fields and by considering the partition function
(also known as generating functional, cf.\ e.g.\ Ref.~\cite{BDJ})
\begin{eqnarray}\label{eq:ZYM}
  {\cal Z}[j,\bar\eta,\eta]&=&\int[dA][d\bar c][dc]\exp\Big(-S_{YM}
  +\int d^4x\frac1{2\xi}(\partial_\mu A^\mu_a)(\partial_\nu A^\nu_a)
  \strut\nonumber\\&&\strut
  +\int d^4x\left(\bar c^a\partial_\mu\partial^\mu c^a
  + g\bar c^af^{abc}\partial_\mu A^{b\mu}c^c\right)
  -\int d^4x(\bar\eta_ac^a+\bar c^a\eta_a)\Big).
\end{eqnarray}
From the partition function, we can derive the pure correlation functions
\begin{eqnarray}
G_{n\mu_1\mu_2\ldots\mu_n}^{a_1a_2\ldots a_n}(x_1,x_2,\ldots,x_n)
  &=&\frac{\delta^n\ln Z[j,\bar\eta,\eta]}{\delta j^{\mu_1}_{a_1}(x_1)
  \delta j^{\mu_2}_{a_2}(x_2)\cdots\delta j^{\mu_n}_{a_n}(x_n)}, \nonumber \\
P_n^{a_1a_2\ldots a_n}(x_1,x_2,\ldots,x_n)
  &=&\frac{\delta^n\ln Z[j,\bar\eta,\eta]}{\delta\bar\eta_{a_1}(x_1)
  \delta\bar\eta_{a_2}(x_2)\cdots\delta\bar\eta_{a_n}(x_n)}, \nonumber \\
{\bar P}_n^{a_1a_2\ldots a_n}(x_1,x_2,\ldots,x_n)
  &=&\frac{\delta^n\ln Z[j,\bar\eta,\eta]}{\delta\eta_{a_1}(x_1)
  \delta\eta_{a_2}(x_2)\cdots\delta\eta_{a_n}(x_n)}
\end{eqnarray}
and mixed correlation functions, like for instance
\begin{equation}
K_{2\mu}^{ab}(x_1,x_2)=\frac{\delta^2\ln Z[j,\bar\eta,\eta]}{\delta
  j^a_\mu(x_1)\bar\eta^b(x_2)}.
\end{equation}
Our aim is to find a Gaussian solution where all the correlation functions can
be expressed by $G_1$ and $G_2$, given the expectation values of the gauge
field, that is, the cumulant expansion of the partition function. It should be
emphasized that the first two Dyson--Schwinger equations give the analogues of
the classical equations of motion and the propagator of the theory,
respectively, if the quantum corrections are properly accounted for.

This aim can be achieved by solving the Dyson--Schwinger
equations~\cite{Frasca:2015yva,Frasca:2016sky,Frasca:2017slg,Chaichian:2018cyv}
\begin{eqnarray}
\label{DSE2}
\lefteqn{\partial^2G_{1\nu}^{a}(x)+gf^{abc}
  \Big(G_{1\mu}^{b}(x)\partial^\mu G_{1\nu}^{c}(x)
  -G_{1\mu}^{b}(x)\partial_\nu G_{1}^{\mu c}(x)
  +\partial^\mu\left(G_{1\mu}^b(x)G_{1\nu}^c(x)\right)\Big)}\nonumber\\&&
  +g^2f^{abc}f^{cde}\Big(G_{2\mu}^{\mu bd}(x,x)G_{1}^{\nu e}(x)
  +G_{2\nu}^{\mu be}(x,x)G_{1\mu}^{d}(x)\nonumber\\&&\kern36pt
  +G_{2\mu\nu}^{de}(x,x)G_{1}^{\mu b}(x)
  +G_{1}^{\mu b}(x)G_{1\mu}^{d}(x)G_{1\nu}^{e}(x)\Big)=0,
\end{eqnarray}
and
\begin{eqnarray}\label{eq:ds_32}
\lefteqn{\partial^2G_{2\nu\kappa}^{am}(x,y)+gf^{abc}\Big(\partial^\mu
  G_{2\mu\kappa}^{bm}(x,y)G_{1\nu}^{c}(x)
  +\partial^\mu G_{1\mu}^{b}(x)G_{2\nu\kappa}^{cm}(x,y)
  -\partial_\nu G_{2\mu\kappa}^{bm}(x,y)G_{1}^{\mu c}(x)} \nonumber\\ && 
  -\partial_\nu G_{1\mu}^{b}(x)G_{2\kappa}^{\mu cm}(x,y)
  +\partial^\mu\Big(G_{2\mu\kappa}^{bm}(x,y)G_{1\nu}^{c}(x)
  +G_{1\mu}^{b}(x)G_{1\nu\kappa}^{cm}(x,y)\Big) \nonumber\\&&\strut\kern-12pt
  +g^2f^{abc}f^{cde}\Big(G_{2\mu\nu}^{bd}(x,x)G_{2\kappa}^{\mu em}(x,y)
  +G_{2\nu\rho}^{eb}(x,x)G_{2\kappa}^{\rho dm}(x,y)
  +G_{2\kappa}^{\mu bm}(x,y)G_{1\mu}^{d}(x)G_{1\nu}^{e}(x) \nonumber \\ &&
  +G_{1}^{\mu b}(x)G_{2\mu\kappa}^{dm}(x,y)G_{1\nu}^{e}(x)
  +G_{1}^{\mu b}(x)G_{1\mu}^{d}(x)G_{2\nu\kappa}^{em}(x,y))
  =\delta_{am}g_{\nu\kappa}\delta^4(x-y).\qquad
\end{eqnarray}
To solve these equations, we use a mapping theorem on the solutions of the
classical Yang--Mills equations of motion~\cite{Frasca:2007uz,Frasca:2009yp}
in order to take all components for the 1P-correlation function as equal,
\begin{equation}
G_{1\mu}^a(x)=\eta_\mu^a\phi(x),
\end{equation}
where $\eta_\mu^a$ are sets of constants (e.g.,
$\eta=((0,1,0,0),(0,0,1,0),(0,0,0,1))$ for $SU(2)$), and $\phi$ is a scalar
field to be determined. We also take
\begin{equation}
  {\tilde G}_{2\mu\nu}^{ab}(x,y)=\delta^{ab}\eta_{\mu\nu}\Delta(x,y)
\end{equation}
where we have made explicit the projector and $\Delta(x,y)$ is a scalar
propagator to be determined. Anticipating this {\it ansatz\/}, the
anti-symmetry of $f^{abc}$ was already used in Eqs.~(\ref{DSE2})
and~(\ref{eq:ds_32}). In doing so, we are able to obtain both $G_1$ and $G_2$
in closed form and, in principle, compute the partition function at any
desired order for this Gaussian solution. By direct substitution, we see
immediately that
\begin{eqnarray}\label{eq:DSphi}
\partial^2\phi+2\lambda\Delta(x,x)\phi+\lambda\phi^3&=&0 \nonumber \\
\partial^2\Delta(x,y)+2\lambda\Delta(x,x)\Delta(x,y)+3\lambda\phi^2(x)
  \Delta(x,y)&=&\delta^4(x-y)\qquad
\end{eqnarray}
with $\lambda=Ng^2$. As $\Delta(x,x)$ is a constant, we can identify this as a
mass correction that arises from quantum effects,
\begin{equation}
\delta m^2=2\lambda\Delta(x,x).
\end{equation}
In the following we will neglect this correction that, after renormalization,
is assumed to have a small effect on the spectrum of the
theory~\cite{Frasca:2017slg}. The two-point correlation function we obtain is
translation invariant. We show this in Appendix~A. Though the description of
the Yang--Mills theory by a Lagrange density invariant unveils a symmetry
under the $SU(N)$ gauge group that is broken by the introduction of the gauge
fixing parameter, this is not the symmetry we concentrate on in this paper.
Instead, we investigate the breaking of the chiral symmetry of this theory by
a mass gap generated dynamically. 

\section{Partition function in the IR regime}
Neglecting the mass correction, the solution for $G_1$ and $G_2$ can be
obtained by taking
\begin{equation}\label{solG1}
G_{1\mu}(x)=\eta_\mu^a\mu\sn\left(p\cdot x+\theta|-1\right)
\end{equation}
in terms of integration constants $\mu$ and $\theta$, where the gluon momentum
$p$ satisfies the dispersion relation $p^2=\mu^2\lambda/2$, and
$\sn(\zeta|\kappa)$ is the Jacobi elliptic function of the first kind with
parameter $\kappa$. In our approach, $G_1$ is treated as a background field or
as a specific non-trivial vacuum. For more details on the solution, cf.\
Appendix~B.

In a first approximation that holds in the infrared (IR) limit of the
Yang--Mills theory (confined phase), our approach allows to truncate the
functional series derived from Eq.~(\ref{eq:ZYM}) at the quadratic term with
the exact translationally invariant two-point function
$G_{2\mu\nu}^{ab}(x_1,x_2)\equiv \tilde{D}_{\mu\nu}^{ab}(x_1-x_2)$, while the
three- and four-point functions are represented as products of one- and
two-point functions evaluated at different points. Because of this, the theory
is manifestly translation invariant at the level of observables where the
Lehmann--Symanzik--Zimmermann reduction formula is assumed to hold. 

In the considered approximation with the partition function truncated at the
quadratic order, the IR truncated partition function
reads~\cite{Frasca:2021mhi}
\begin{equation}\label{Z_YM-IR}
Z_{\rm YM}[j]\Big|_{\rm IR}\approx
Z_{\rm YM}[0]\Big|_{\rm IR}\,\exp\Big\{-\int d^4xG_{1\mu}^a(x)j^{a\mu}(x)
-\frac{1}{2}\int d^4x d^4yj^{a\mu}(x)\tilde G_{2\mu\nu}^{ab}(x-y)j^{b\nu}(y)
\Big\}.
\end{equation}
The momentum space propagator in Feynman gauge takes the shape
\begin{equation}\label{eq:gprop}
G_{2\mu\nu}^{ab}(p)=\delta_{ab}\eta_{\mu\nu}\sum_{n=0}^\infty
  \frac{B_n}{p^2-m_n^2+i\epsilon}
\end{equation}
(see Appendix~A for a derivation), where
\begin{equation}\label{eq:Bn}
B_n=(2n+1)^2\frac{\pi^3}{4K^3(-1)}
  \frac{e^{-\left(n+\frac{1}{2}\right)\pi}}{1+e^{-(2n+1)\pi}},
\end{equation}
and the mass spectrum reads
\begin{eqnarray}\label{mass-spectrum}
m_n=(2n+1)m_0,\qquad
m_0\equiv\frac{\pi\mu}{2K(-1)},
\end{eqnarray}
in terms of the mass of the lowest excitation, $m_0$. Note that $m_0$ is a
physical parameter in the same way as the scale of asymptotic freedom is
generated dynamically in QCD by perturbation theory. We can identify this
parameter with the string tension normally evaluated at $\sigma=(440\MeV)^2$.
Stopping to the two-point correlation function, one obtains
\begin{eqnarray}
  Z_{\rm YM}[j]&\propto&\exp\left[-\frac12\int d^4pj^\mu_a(p)
    G_{2\mu\nu}^{ab}(p)j^\nu_b(-p)\right] \nonumber \\
  &=&\exp\left[-\frac12\int d^4pj_\mu^a(p)\sum_{n=0}^\infty
    \frac{B_n}{p^2-m_n^2+i\epsilon}j^\mu_a(-p)\right].
\end{eqnarray}
Due to the sum, this is equivalent to a product of partition functions of an
infinite set of scalar fields $\{\phi_n\}$ (i.e.\ glue states of mass $m_n$),
each with weight $n_g=2(N^2-1)$ (the factor 2 arises from the Minkowski metric,
assuming all the currents are equal). Such a truncation represents a gas of
free massive glue states. For such a gas we expect a very good agreement at
low temperatures, as we will see below. Finally, one can use the functional
identity 
\begin{eqnarray}
\lefteqn{\left\{\exp\left[-\frac{1}{2}\int d^4pj(p)\sum_{k=0}^\infty
    \frac{B_k}{p^2+m_k^2}j(-p)\right]\right\}^{n_g}}\nonumber \\
  &=&\left\{\prod_{k=0}^\infty\int{\cal D}\phi_k(p)
  \exp\left[\frac{-1}{2B_k}\int d^4p\phi_k(p)(p^2+m_k^2)\phi_k(-p)
  -\int d^4pj(p)\phi_k(-p)\right]\right\}^{n_g}.\qquad
\end{eqnarray}
Turning back to the partition function of each of the scalar field separately,
one can write the Yang--Mills partition function in Eq.~(\ref{Z_YM-IR}) as
\begin{equation}\label{eq:ZI}
Z_{\rm YM}[0]\Big|_{\rm IR} \propto Z_{\rm IR} \equiv
  \Big[\prod_k \int {\cal D}\phi_k \, \exp\Big\{-\frac{1}{2B_k}
  \int\frac{d^4p}{(2\pi)^4}\phi_k(p)(p^2+m_k^2)\phi_k(-p)\Big\}\Big]^{n_g} 
\end{equation}
up to an overall multiplicative constant that is irrelevant for further
considerations. Indeed, the source-free Yang--Mills partition function in the
non-perturbative regime can be seen as a partition function of a system of
infinitely many free scalar fields, where each field contributes with
different weights, depending on its mass $m_k$ determined by Eq.~(\ref{eq:ZI}). 

We see that, in the IR limit, the spectrum of the theory is quite different
from that of the UV limit where the true states of the theory are massless
gluons and we have asymptotic freedom. In the IR limit, we have an infinite
spectrum of massive particles that, in our approximation, are not interacting
with each other. If we use this spectrum for the regime of asymptotic freedom
as well, i.e.\ for high temperatures, this will yield an infinite free energy
for the model. We will see that for the high-temperature limit, the omission
of the gluon states beyond the lowest one is a fairly good approximation.

\section{Thermal Yang--Mills theory}
In what follows, we aim to the basderive ic thermodynamic properties of the
thermal Yang--Mills theory. Note that Eq.~(\ref{eq:ZI}) is the partition
function of an infinite number of free scalar fields. Raised to the exponent,
the product over the scalar fields adds a further sum over the scalar field
modes. In the imaginary-time formulation, Eq.~(\ref{eq:ZI}) can be rewritten
as~\cite{Bellac}
\begin{eqnarray}
  \ln Z_{\rm IR} = -\frac{n_g}{2}\, \sum_k\sum_{n=-\infty}^\infty\int\frac{d^3p}
      {(2\pi)^3}\ln\left(\beta^2\,B_k^{-1}\,(\omega_n^2+{\bm p}^2+m_k^2)\right),
\end{eqnarray}
where $\omega_n=2n\pi/\beta$ (with $\beta\equiv 1/T$) are the Matsubara
frequencies. Furthermore, making use of the representation
\begin{eqnarray} \nonumber
\ln[(2\pi n)^2+\beta^2\epsilon_k^2]=\int_1^{\beta^2\epsilon_k^2}
  \frac{dx^2}{x^2+(2\pi n)^2}+\ln[1+(2\pi n)^2]
\end{eqnarray}
with $\epsilon_k^2 = \epsilon_k^2({\bm p}) \equiv {\bm p}^2+m_k^2$, performing
the summation
\begin{eqnarray}
\sum_{n=-\infty}^\infty\frac{1}{n^2+(x/2\pi)^2}=\frac{2\pi^2}{x}
\left(1+\frac{2}{e^x-1}\right),
\end{eqnarray}
and turning to the continuum limit for momentum $\bm p$ in the system of
volume $V$, we arrive at the final form for the partition function of the
thermal Yang--Mills theory, given by
\begin{equation}\label{ZIR2}
\ln Z_{\rm IR} = n_g V\, \sum_{k=0}^\infty\int\frac{d^3p}{(2\pi)^3}\Big[
  -\frac{\beta\epsilon_k({\bm p})}2-\ln\left(1-e^{-\beta\epsilon_k({\bm p})}
  \right)\Big].
\end{equation}
Here, the first expression in square brackets represents the contribution of
the zero-point fluctuations of fields with mass $m_k$ in the ground state. In
general, the vacuum term $\ln Z_{\rm IR}^{\rm vac}$ does not affect
thermodynamic observables in the thermal equilibrium but effectively
renormalizes the physical observables by absorbing divergences emerging in the
summation over $k$ as $T\to \infty$, such that the physics is recovered
through a finite, properly regulated, partition function, $\ln Z_{\rm IR}
\to\ln Z^{\rm reg}_{\rm R}\equiv\ln Z_{\rm IR}-\ln Z_{\rm IR}^{\rm vac}$.

For further considerations, it is instructive to turn to a more compact
representation in terms of a dimensionless variable $z=p\beta$ (with
$p\equiv |{\bm p}|$), such that
\begin{eqnarray}
\ln Z^{\rm reg}_{\rm R}&=&-P_0V\beta g(\alpha),\qquad
  P_0=\frac{n_g}{\beta^4}, \\
g(\alpha)&\myeq&\sum_{k=0}^\infty J(a_k),\quad
  a_k\equiv(2k+1)\alpha=\frac{m_k}{T}, \\
J(a_k)&\equiv&\frac1{2\pi^2}\int_0^\infty z^2dz
  \ln\left(1-e^{-\sqrt{z^2+a_k^2}}\right),
\label{J_B}
\end{eqnarray}
where a proper regularization of the summation is implied, $J(a_k)$ is
the standard thermal integral for a given state with mass $m_k$ defined in
Eq.~(\ref{mass-spectrum}), $\alpha\equiv\beta m_0$ is the inverse of the
temperature measured in units of the mass gap $m_0$, and $P_0$ is the pressure
of the free gluon gas. In what follows, we are interested in analyzing such
thermodynamic observables of the quantum Yang--Mills fields like pressure
\begin{eqnarray}
P=\beta^{-1}\frac{\partial}{\partial V}\ln Z^{\rm reg}_{\rm R}=-P_0g(\alpha),
\end{eqnarray}
that is essentially the (normalized) partition function (cf.\
Eq.~(\ref{ZIR2})), and energy density
\begin{eqnarray}
\epsilon=-\frac1V\frac\partial{\partial\beta}\ln Z^{\rm reg}_{\rm R}
  =\frac\partial{\partial\beta}\left(\beta g(\alpha)\right)
\end{eqnarray}
as well as the trace of the energy-momentum tensor (trace anomaly),
$\Delta\equiv\epsilon-3P$. 

The characteristic energy scale for the Yang--Mills theory will enter in our
analysis through the ratio $m_0/T_c$, where $m_0$ is the mass gap of the
theory and $T_c$ the critical temperature of the phase transition, that will
be the only parameter to be fitted against the lattice data. The critical
temperature is known from lattice data being around $271\MeV$ for $SU(3)$ and
around $262\MeV$ for $SU(4)$ according to Ref.~\cite{Lucini:2012gg}. 

\section{Evaluation of the partition function}
In order to determine pressure and energy density, we have to evaluate
$g(\alpha)$. This is known in a form of a series as
\begin{equation}\label{eq:gser}
g(\alpha)=-\frac{\alpha^2}{2\pi^2}\sum_{k=0}^\infty\sum_{n=1}^\infty
  \frac1{n^2}(2k+1)^2K_2(n(2k+1)\alpha)
\end{equation}
(see Appendix~C for a derivation), where $K_2(x)$ is a modified Bessel
function. In the low-temperature limit, no convergence problems arise, and the
series can be trusted enough for a comparison with lattice data. On the other
hand, for $\alpha\rightarrow 0$, i.e., the high-temperature limit, this is no
longer true for the series~(\ref{eq:gser}), as it should be expected by the
infinite spectrum of free particles we are considering. The reason is that we
are using an IR approximate solution to evaluate the UV limit of the
Yang--Mills theory. This would be possible only by setting the mass gap of the
theory to zero. 

\begin{figure}
\centering
\includegraphics[height=8cm]{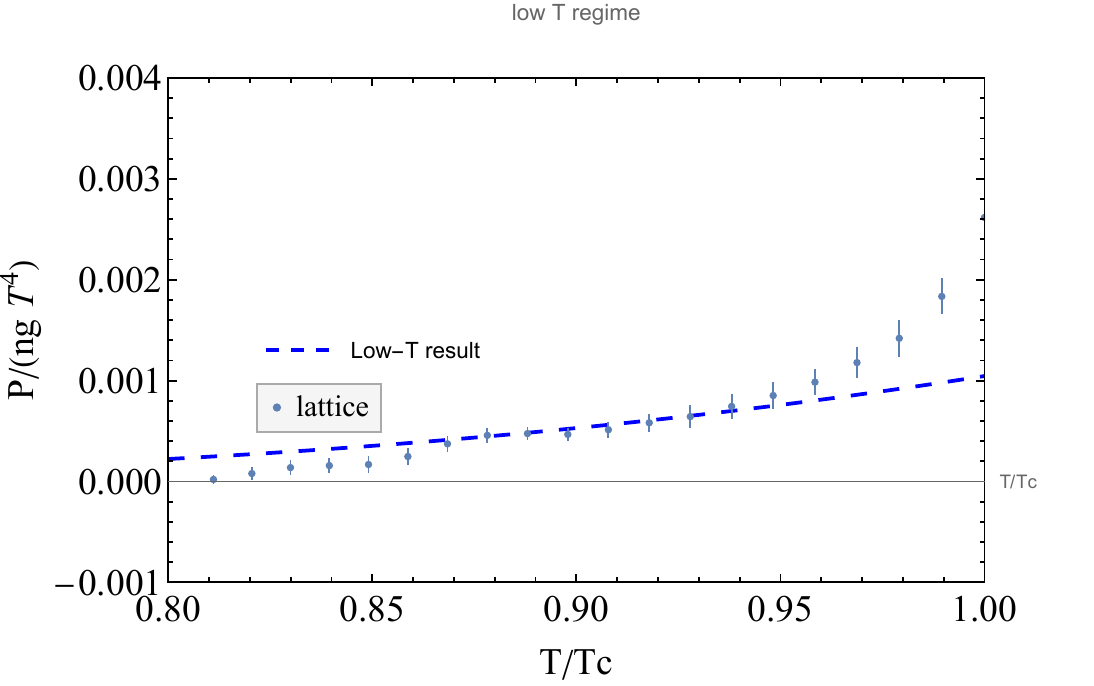}\\[12pt]
\includegraphics[height=8cm]{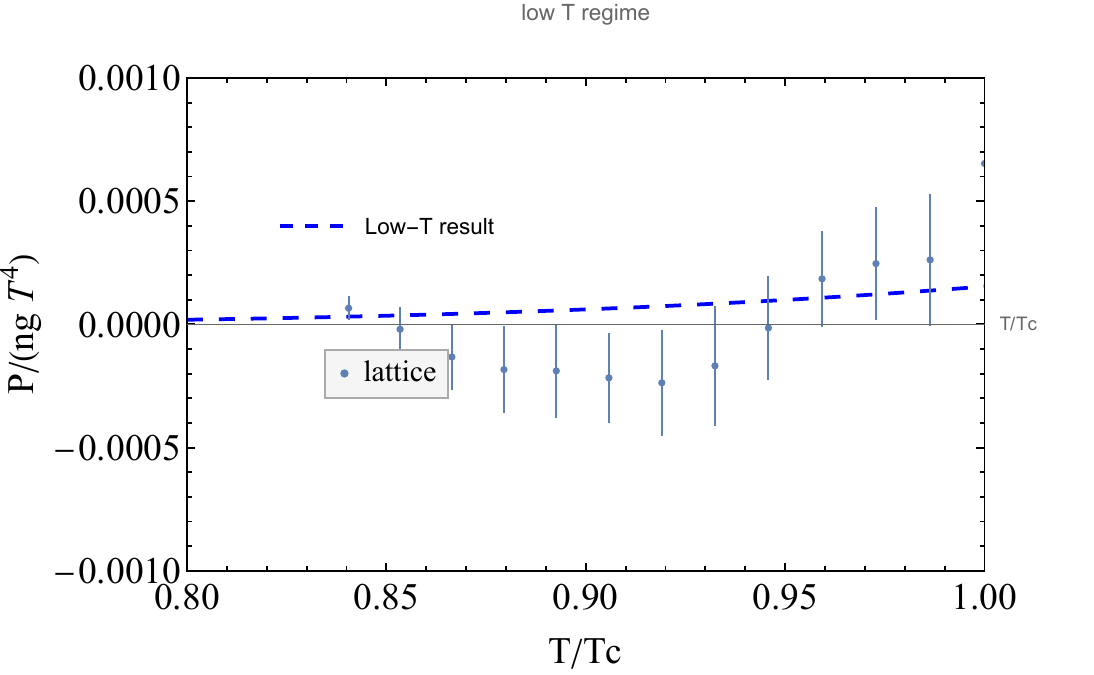}
\caption{\it \textbf{Upper:} Plot of the pressure $p$ for $SU(3)$ with the
 fitted ratio $m_0/T_c\approx 1.61$ compared to the lattice data.
\textbf{Lower:} The same for $SU(4)$ with the fitted ratio
$m_0/T_c\approx 1.95$.\label{fig1}}
\end{figure}

\begin{figure}
\centering
\includegraphics[height=8cm]{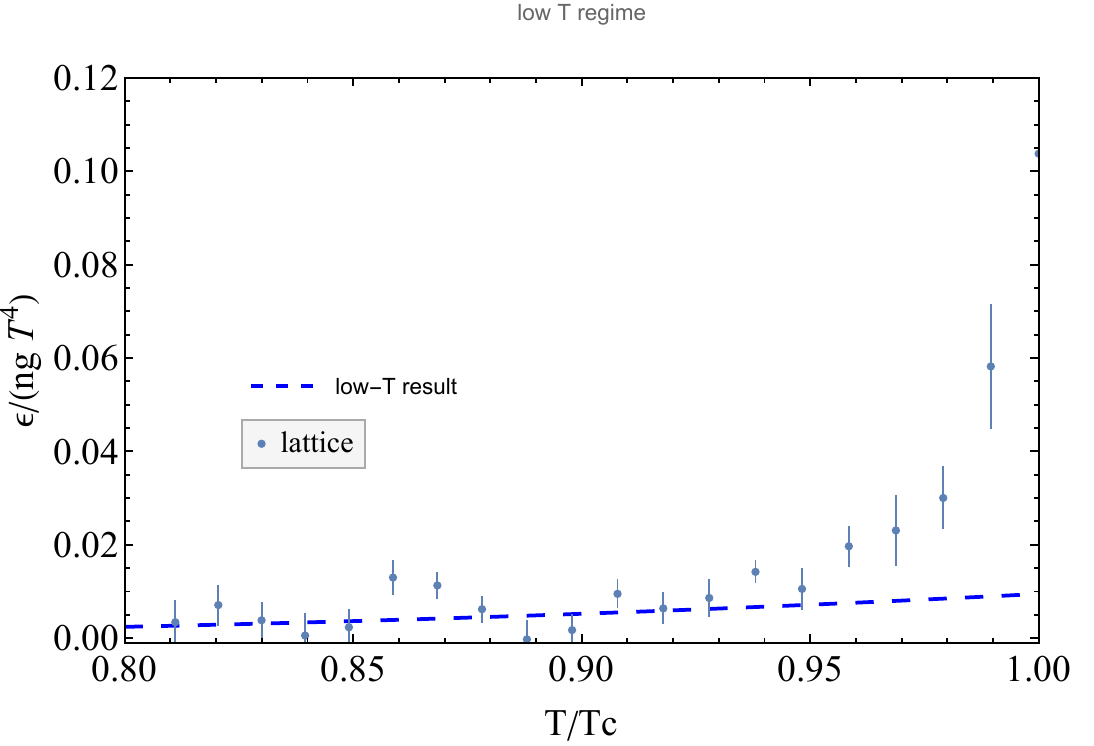}\\[12pt]
\includegraphics[height=8cm]{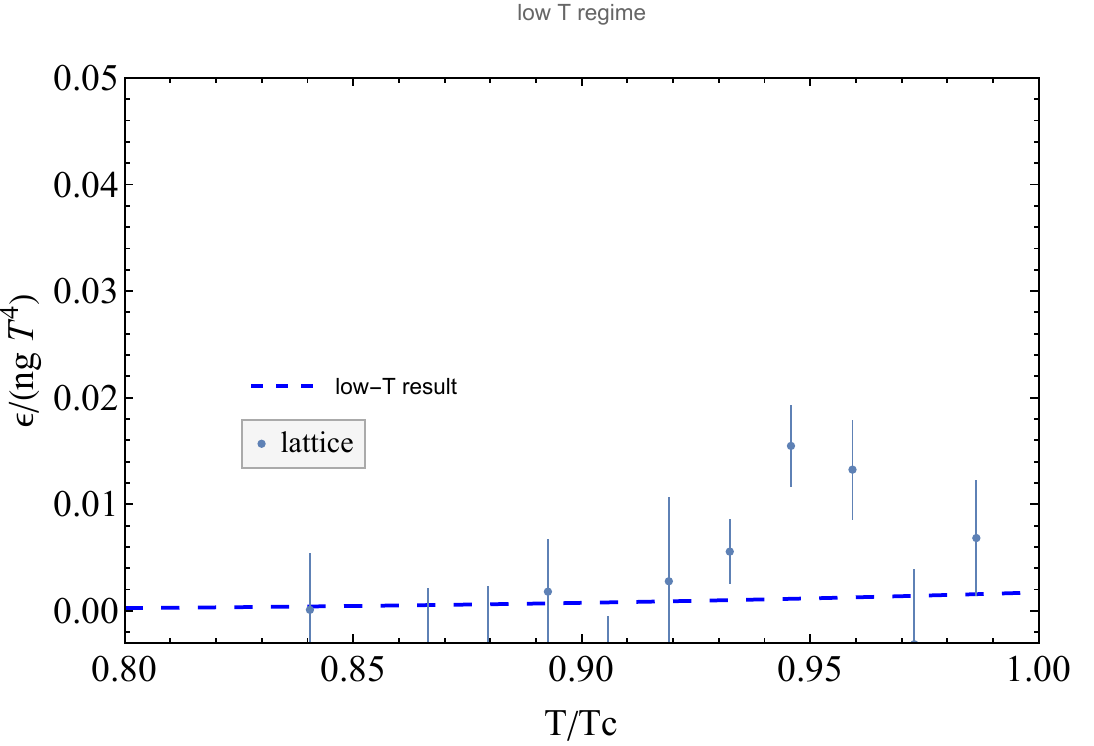}
\caption{\it \textbf{Upper:} Plot of the energy density $\epsilon$ for $SU(3)$
with the fitted ratio $m_0/T_c\approx 1.61$ compared to the lattice data.
\textbf{Lower:} The same for $SU(4)$ with the fitted ratio
$m_0/T_c\approx 1.95$.\label{fig2}}
\end{figure}

In the following plots we compare our results for the pressure and the energy
density for $SU(3)$ and $SU(4)$ with those from lattice. As a single fitting
parameter for both plots we use the ratio $m_0/T_c$. For the pressure
displayed in Fig.~\ref{fig1} we see a fairly good approximation
$m'_0/T_c\approx (m_0/T_c)(N/3)^{-1/4}$ where $m_0'$ is the mass gap for
$SU(3)$. Both masses are obtained independently by a fit. Because of this,
they already take into account the number of colors. For the energy density
we obtain the result shown in Fig.~\ref{fig2}. Again we have good agreement,
with the dependency on the number of colors the same as for the pressure. We
see that the ratio $m_0/T_c$, both for pressure and energy density, is very
similar within the errors of the corresponding fits.

\begin{figure}
\centering
\includegraphics[height=8cm]{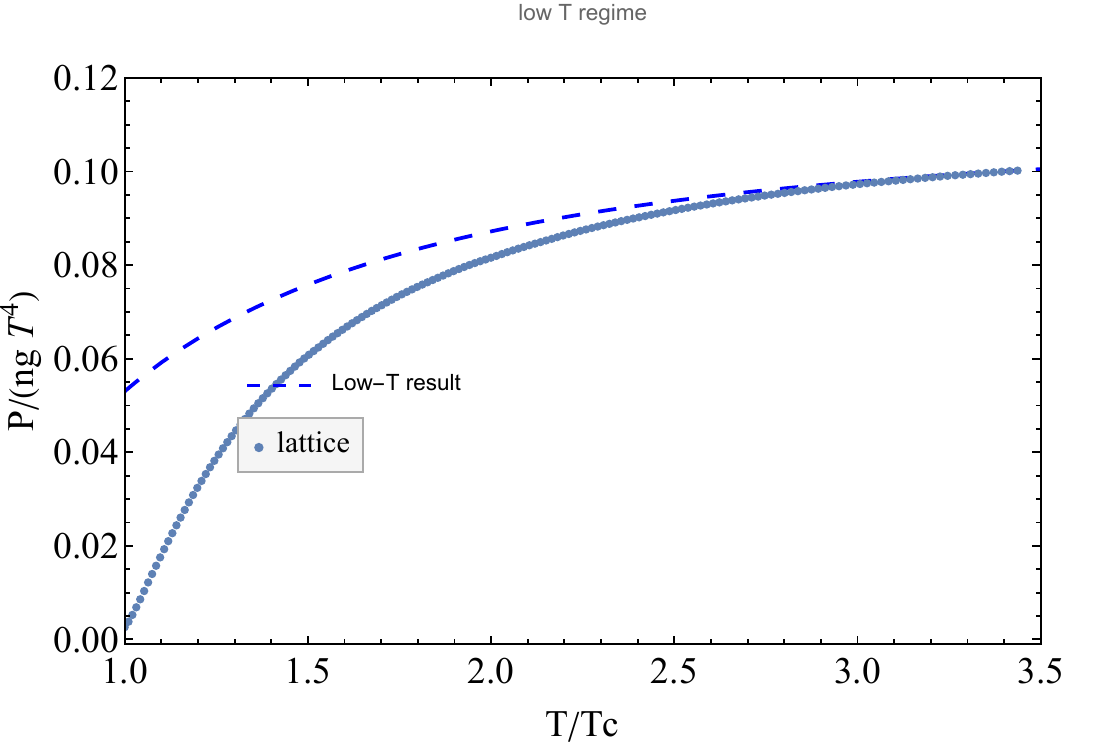}\\[12pt]
\includegraphics[height=8cm]{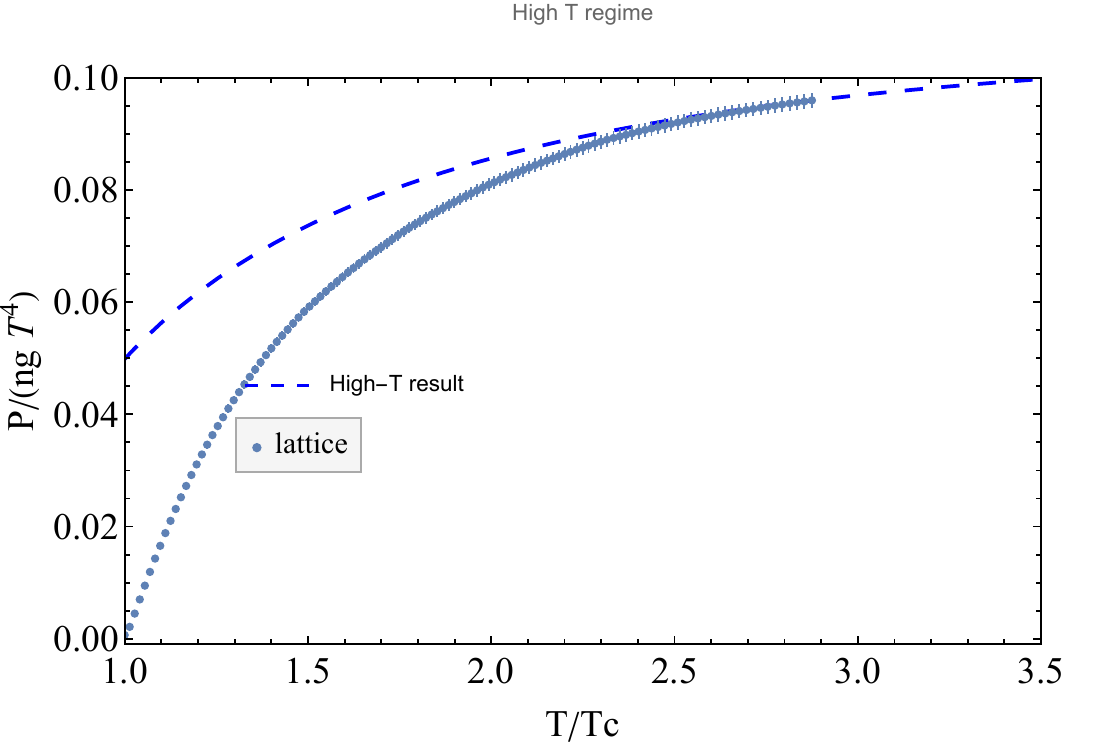}
\caption{\it \textbf{Upper:} Plot of the pressure $p$ for $SU(3)$ in the
high-temperature limit with the fitted ratio $m_0/T_c\approx 0.53$ compared to
the lattice data. \textbf{Lower:} The same for $SU(4)$ with the fitted ratio
$m_0/T_c\approx 0.41$. \label{fig3}}
\end{figure}

\begin{figure}
\centering
\includegraphics[height=8cm]{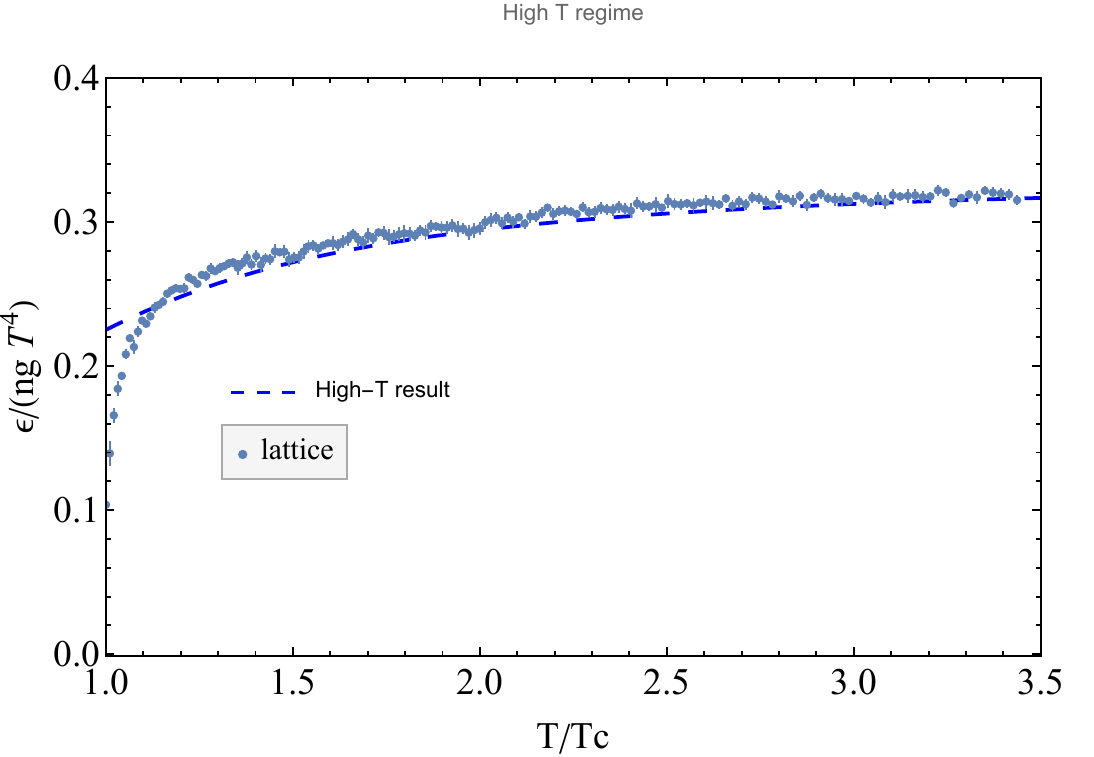}\\[12pt]
\includegraphics[height=8cm]{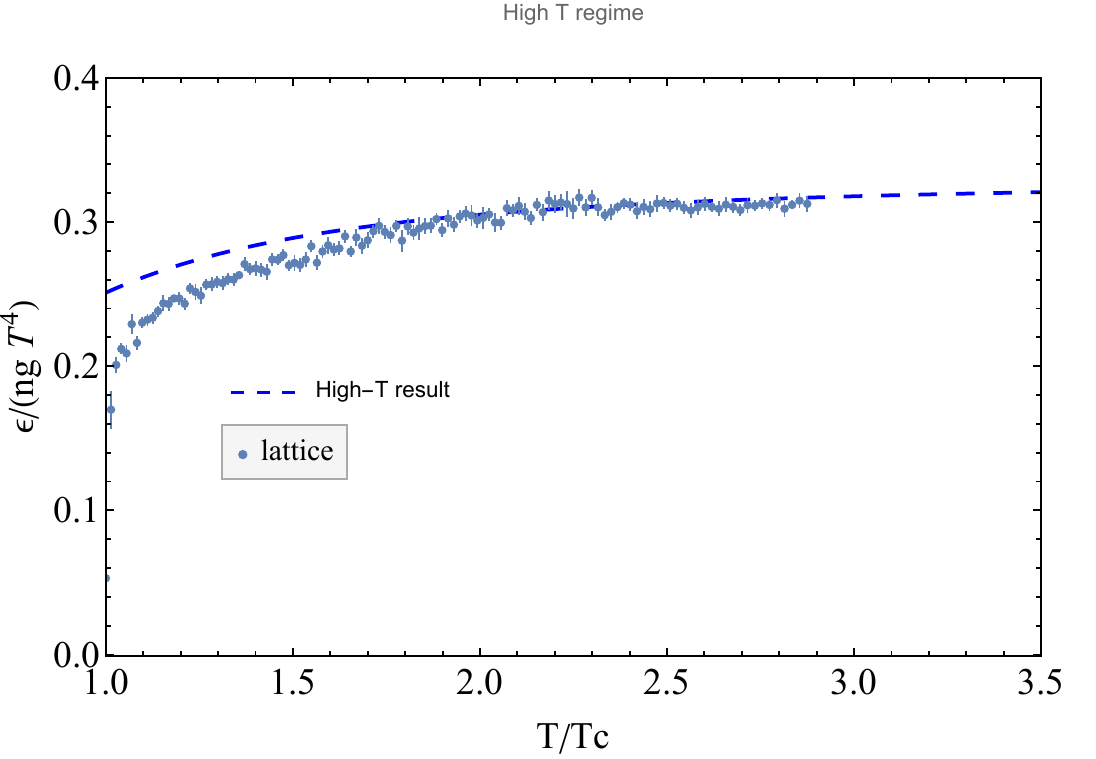}
\caption{\it \textbf{Upper:} Plot of the energy density $\epsilon$ for $SU(3)$
with the fitted ratio $m_0/T_c\approx 0.53$, as compared to the lattice data.
\textbf{Right:} The same for $SU(4)$ with the fitted ratio
$m_0/T_c\approx 1.59$.\label{fig4}}
\end{figure}

In the high-temperature regime where perturbative QCD can be satisfactorily
applied~\cite{Borsanyi:2012ve}, we achieve the somewhat surprising result that
the corresponding thermal series for a scalar field seems to fit very well the
results, mostly in the case of the energy density. We would like to emphasize
that the fits we obtain in this regime cannot be successful for the same
values of the ratio $m_0/T_c$. The reason is that we are just extending our
approach with a spectrum of massive glue particles to a regime with massless
gluons. In addition, the glueball mass can entail some dependence on the
temperature that we are not able to catch at this stage~\cite{Silva:2013maa}.
The achievement of satisfactory results also in the high-temperature limit
(i.e., for small values of $\alpha$) will grant some extended discussion. For
this we consider the expansion
\begin{eqnarray}
g(\alpha)&=&\frac{\pi^2}{90}-\frac{\alpha^2}{24}+\frac{\alpha^3}{12\pi}
  +\frac{\alpha^4}{64\pi^2}\left(\log(\alpha^2)-\frac{3}{2}-2\log(4\pi)
  +2\gamma_E\right)\nonumber \\ && \strut
  +\pi^\frac{3}{2}\sum_{k=1}^\infty(-1)^k\frac{\zeta(2k+1)}{(k+1)!}
  \Gamma\left(k+\frac{1}{2}\right)\left(\frac{\alpha^2}{4\pi^2}\right)^{k+2},
\end{eqnarray}
where $\gamma_E$ is the Euler--Mascheroni constant. This corresponds to
considering Eq.~(\ref{eq:gser}) just for a single term and expanding for
high temperatures for $\alpha\rightarrow 0$. The corresponding plots are
given in Figs.~\ref{fig3} and~\ref{fig4}. From these plots it can be easily
observed how harsh our approximation is compared to a standard perturbative
computation. Nevertheless, for the energy density in the high-temperature
regime we obtain Fig.~\ref{fig4}. The agreement in this case is excellent with
the rule for the ratio $m_0/T_c$ and the number of colors satisfied again.

We have fitted each quantity separately, as we cannot expect that the ratio
$m_0/T_c$ will stay the same, and for consistency reasons. For the
high-temperature limit, the ratio $m_0/T_c$ for the energy density is
somewhat distant from the one of the pressure for $SU(4)$. Overall, the
agreement is excellent for a single fit parameter. Taking into account that we
are working with asymptotic series with zero radius of convergence, we seem to
have evaluated the optimal number of terms for a satisfactory agreement with
lattice data.

\section{Discussion and conclusions}
In this work, we have presented the non-perturbative partition function of the
thermal $SU(N)$ Yang--Mills theory in the framework of Dyson--Schwinger
equations with a non-trivial ground state in analytic form. Utilizing a novel
technique for solving these equation in a differential form, we find very
good consistency of the resulting thermodynamical observables -- pressure and
energy density -- with the corresponding lattice results available for $N=3$
and $N=4$ colors. Our analytical approach provides a pathway to obtain
thermodynamic characteristics in terms of input parameters to a given action
of the Yang--Mills theory. Indeed, in the considered IR limit of the theory,
we show that the functional series for the partition function, already
truncated at the quadratic term, provides good agreement between our results
and the lattice data at low temperatures. While a standard technique could be
applied for high temperatures, we considered the solution for the scalar field
and found very good agreement for the energy density.

As the quadratic approximation already captures the most essential features of
non-perturbative Yang--Mills dynamics at low temperatures, we believe that,
with a complete evaluation of the partition function going beyond the
quadratic order and accounting for higher-order correlation functions, a more
precise behavior could be found, possibly with evidence for a smooth
transition from a low-energy spectrum with glue states to that of a pure
massless gluon spectrum, as observed in the asymptotic freedom regime.

A scenario with a spectrum of massive glue state excitations in a thermal
setting for the Yang--Mills theory, starting from the available lattice data,
has already been devised in literature~\cite{Peshier:1995ty,Castorina:2011ra}.
In our study, we demonstrate that the same behavior can be obtained
analytically.

Cosmological first-order phase transitions (FOPT) in the early universe are
known to radiate gravitational waves (GWs) which are potentially detectable in
upcoming GW missions like LISA~\cite{LISACosmologyWorkingGroup:2022jok,%
LISA:2017pwj,Janssen:2014dka,Yagi:2011wg}. If such a signal would be detected,
this would inevitable give hints for new physics, since the electroweak and
QCD phase transitions in the standard model are known as cross-overs, not as
FOPT. Consequently, studying such GW signals will allow us to probe the
dynamics of otherwise completely inaccessible dark or hidden
sectors~\cite{Schwaller:2015tja,Breitbach:2018ddu,Fairbairn:2019xog}.
Therefore, we envisage a possible application of our approach to studies of
such phase transitions in the early universe, in particular for the strong
coupling. $SU(N)$ Yang--Mills theories featuring a color confinement string
FOPTs~\cite{Svetitsky:1982gs,Panero:2009tv}, naturally appear in several
well-motivated viable extensions of the standard model, see studies in
Refs.~\cite{Gross:1984dd,Dixon:1985jw,Dixon:1986jc,Acharya:1998pm,%
Halverson:2015vta,Asadi:2021pwo,Asadi:2021yml,Bai:2013xga,Schwaller:2015gea}.
Since the only free and independent parameter here is the confinement scale,
these scenarios can be deemed minimal, thus serving as suitable benchmark
models. Previous attempts to study the GW signal in such models in
Refs.~\cite{Halverson:2020xpg,Bigazzi:2020phm,Bigazzi:2020avc,%
Huang:2020mso,Wang:2020zlf,Kang:2021epo,Yamada:2022aax,Yamada:2022imq} suffer
from uncertainties related to non-perturbative effects, or rely on lattice
calculations or methods like AdS/CFT~\cite{Morgante:2022zvc} due to the strong
coupling involved. We envisage the method developed in this paper to make
quantitative estimates for such scenarios. As this is beyond the scope of the
present paper, we plan to take up this subject in a future publication, given
the timeline of the LISA experiment upcoming very soon and LiGO taking data
already now.

Our results are expected to have a strong impact also on a possible resolution
of the long-standing problem of deriving the QCD equation of state and the
precise reconstruction of the QCD phase diagram, critical for many ongoing
studies in particle physics (e.g.\ physics of the quark--gluon plasma) and
astrophysics (e.g.\ the dynamics of neutron stars), and other issues in
cosmology (e.g.\ dark matter, dark energy and inflation), see
Refs.~\cite{Hindmarsh:2005ix,Drees:2015exa,Hajkarim:2019csy}.

\section*{Acknowledgements}
The authors thank Marco Panero for providing the data of his lattice
computations for purpose of comparison. We are also grateful to Roman
Pasechnik and Zhi-Wei Wang who were greatly helpful during the first stage of
this project.

\begin{appendix}
\section{Two-point correlation function and spectrum}
\setcounter{equation}{0}\def\theequation{A\arabic{equation}}  
We have to solve the set of Dyson--Schwinger equations (\ref{eq:DSphi}), cf.\
Refs.~\cite{Frasca:2015yva,Frasca:2016sky,Frasca:2017slg,Chaichian:2018cyv}.
The aim is to obtain a Fubini--Lipatov solution to represent the vacuum of the
theory and a translation invariant two-point correlation function. Actually,
both the equations can be solved analytically using Jacobi elliptic
functions~\cite{WW:1927,Bate:1953}. In physics, elliptic integrals and
functions appear for instance for the exact solution of the physical pendulum,
in the small angle approximation limited by the mathematical pendulum that
obeys a linear differential equation. Without going too much into details, the
period of a physical pendulum is expressed in terms of the incomplete
elliptic integral
\begin{equation}
F(\varphi|\kappa)=\int_0^\varphi\frac{du}{\sqrt{1-\kappa\sin^2u}}
\end{equation}
with parameter $\kappa$, where the complete elliptic integral given by
$K(\kappa)=F(\pi/2|\kappa)$. The Jacobi elliptic functions are defined via
this incomplete elliptic integral by inversion of the equation
$z=F(\varphi|\kappa)$, in the sense that $\sn(z|\kappa):=\sin\varphi$,
$\cn(z|\kappa):=\cos\varphi$, and
$\dn(z|\kappa):=(1-\kappa\sin^2\varphi)^{1/2}$. Using the Jacobi elliptic
sine, we can write
\begin{equation}
\phi(x)=\mu\sn\left(p\cdot x+\theta,\kappa\right),\qquad
  \kappa=\frac{\delta m^2-p^2}{p^2},
\end{equation}
where $\delta m^2=2\lambda\Delta(x,x)$, and $\theta$ and $\mu$ are two
integration constants. This function is a solution for the nonlinear
differential equation shown in the first line of~(\ref{eq:DSphi}), provided
the dispersion relation
\begin{equation}
   p^2={\delta m}^2+\mu^2\frac\lambda2
\end{equation}
holds. It is interesting to point out that also $-\phi(x)$ solves the first
equation in~(\ref{eq:DSphi}). Therefore, to choose one or the other solution
spontaneously breaks the $Z_2$ symmetry. Similarly, one can show that the
two-point correlation function can be written in the form 
\begin{equation}\label{eq:pp}
\Delta(p)=\frac{\pi^3}{4K^3(\kappa)}\sum_{n=0}^\infty (-1)^n(2n+1)^2
  \frac{q^{n+1/2}}{1-q^{2n+1}}\frac1{p^2-m_n^2+i\epsilon},
\end{equation}
where $q=\exp\left(-\pi K(1-\kappa)/K(\kappa)\right)$, with $K(\kappa)$ the
complete elliptic integral of the first kind. The technique is shown in
Ref.~\cite{Frasca:2023uaw} for the classical case but also applies
straightforwardly here. The mass spectrum is
\begin{equation}
m_n=(2n+1)\frac{\pi\mu}{2K(\kappa)}.
\end{equation}
In order to get back to Eq.~(\ref{eq:Bn}), we set $\delta m=0$. In this case
we obtain $\kappa=-1$ and $q=\exp\left(-\pi(1-i)\right)$. Inserting this back
into Eq.~(\ref{eq:pp}), we are able to obtain the coefficients given in the
main text.

\section{One-point correlation function}
\setcounter{equation}{0}\def\theequation{B\arabic{equation}}
In taking $A_\mu^a(x)$ instead of $G_{1\mu}^a(x)$, Eq.~(\ref{DSE2}) for
$G_{1\nu}^a(x)$ maps very well the classical equation of motion, given by
\begin{eqnarray}
\lefteqn{\partial^\mu(\partial_\mu A_\nu^{a(0)}-\partial_\nu A_\mu^{a(0)}
  +gf^{abc}A_\mu^{b(0)}A_\nu^{c(0)})}\nonumber\\&&\strut
  +gf^{abc}A^{b(0)\mu}(\partial_\mu A_\nu^{c(0)}
  -\partial_\nu A_\mu^{c(0)}+gf^{cde}A_\mu^{d(0)}A_\nu^{e(0)})=0,\qquad
\end{eqnarray}
once we select the Feynman gauge~\cite{Frasca:2015yva}. However, there is an
important difference: Eq.~(\ref{DSE2}) contains a quantum correction of the
form $G_{2\mu\nu}^{ab}(x,x)$. After regularizing the divergences, this
correction can have the effect of a small shift in the spectrum of the
theory~\cite{Frasca:2017slg} that can be neglected. Our aim is to show how
the Jacobi elliptic functions are suited to this equation. By taking the
mixed symbols, in the simple case of $SU(2)$ ($a=1,2,3$ is the group index),
in the form
\begin{equation}\label{eq:msym}
\eta_\mu^1=(0,1,0,0), \quad \eta_\mu^2=(0,0,1,0), \quad \eta_\mu^3=(0,0,0,1),
\end{equation}
and using the mapping theorem between a scalar field and the Yang--Mills
field~\cite{Frasca:2007uz,Frasca:2009yp},
\begin{equation}
A_\mu^{a(0)}(x)=\eta_\mu^a\phi_0(x),
\end{equation}
$\phi_0(x)$ solves
\begin{equation}\label{eq:eqphi}
\partial^2\phi_0+2g^2\phi_0^3=0,
\end{equation}
that is the first of Eqs.~(\ref{eq:DSphi}) for $SU(2)$ when the quantum
corrections are neglected. Considering a solution of the form
$\phi_0(x)=\mu\sn(p\cdot x+\theta|-1)$, we obtain
\begin{eqnarray}
\partial_t^2\phi_0(x)&=&\mu p_0^2(\cn^2(p\cdot x+\theta|-1)
  -\dn^2(p\cdot x+\theta|-1))\sn(p\cdot x+\theta|-1), \nonumber \\
\nabla^2\phi_0(x)&=&\mu|{\bm p}|^2(\cn^2(p\cdot x+\theta|-1)
  -\dn^2(p\cdot x+\theta|-1))\sn(p\cdot x+\theta|-1).
\end{eqnarray}
Therefore,
\begin{eqnarray}
\lefteqn{\partial^2\phi_0+2g^2\phi_0^3\ =\ \mu\sn(p\cdot x+\theta|-1)
  \times\strut}\\&& 
  \left[(E^2-|{\bm p}|^2)(\cn^2(p\cdot x+\theta|-1)-\dn^2(p\cdot x+\theta|-1))
    +2g^2\mu^2\sn^2(p\cdot x+\theta|-1)\right].\nonumber
\end{eqnarray}
Using the identity $\cn^2(z|-1)-\dn^2(z|-1)=-2\sn^2(z|-1)$~\cite{NIST},
Eq.~(\ref{eq:eqphi}) is solved, provided the dispersion relation
$p^2=\mu^2g^2$ for $SU(2)$ holds. If the quantum corrections are neglected, we
get back
\begin{equation}
G_{1\mu}^a(x)=\eta_\mu^a\phi_0(x)=\eta_\mu^a\mu\sn(p\cdot x+\theta|-1),
\end{equation}
that is Eq.~(\ref{solG1}).

\section{Evaluation of the partition function}
\setcounter{equation}{0}\def\theequation{C\arabic{equation}}
Noting that the exponential function grants convergence for $z>0$, one can
expand
\begin{equation}
\ln\left(1-e^{-\sqrt{z^2+a_k^2}}\right)=-\frac1ne^{-n\sqrt{z^2+a_k^2}}.
\end{equation}
Because of the convergence of this series, one can freely interchange the
integration and summation to obtain
\begin{equation}
J_k=\int_0^\infty\ln\left(1-e^{-\sqrt{z^2+a_k^2}}\right)z^2dz
    =-\sum_{n=1}^\infty\frac{1}{n}\int_0^\infty e^{-n\sqrt{z^2+a_k^2}}z^2dz.
\end{equation}
The integral term can be rewritten in a known form via the change of variables
$w^2=z^2+a_k^2$. This yields (see Ref.~\cite{TheTables}, 3.389.4)
\begin{equation}
\int_{a_k}^\infty w(w^2-a_k^2)^\frac12e^{-nw}dw=\frac1na_k^2K_2(n a_k),
\end{equation}
where $K_2(x)$ is a modified Bessel function of the second kind. This yields
Eq.~(\ref{eq:gser}) in the main text. In physics, Bessel functions describe
the vibration modes of a circular membrane, being solutions to the Bessel
differential equation $x^2f''(x)+xf'(x)+(x^2-\alpha^2)f(x)=0$. Modified Bessel
functions are those that obey the modified Bessel differential equation
$x^2f''(x)+xf'(x)+(x^2+\alpha^2)f(x)=0$. An integral representation of the
modified Bessel function of the second kind (diverging at the origin) is given
by
\begin{equation}
K_\alpha(x)=\int_0^\infty e^{-x\cosh t}\cosh(\alpha t)dt,\qquad
\cosh(z)=\frac12(e^z+e^{-z}).
\end{equation}

\end{appendix}


\begin{thebibliography}{00}

\bibitem{Pisarski:2000eq}
  R.~D.~Pisarski,
  Phys.\ Rev.\ D \textbf{62}, 111501 (2000)

\bibitem{Sannino:2002wb}
  F.~Sannino,
  Phys.\ Rev.\ D \textbf{66}, 034013 (2002)

\bibitem{Panero:2009tv}
  M.~Panero,
  Phys.\ Rev.\ Lett.\ \textbf{103}, 232001 (2009)

\bibitem{Carenza:2022pjd}
  P.~Carenza, R.~Pasechnik, G.~Salinas and Z.~W.~Wang,\\
  Phys.\ Rev.\ Lett.\ \textbf{129}, no.26, 26 (2022)

\bibitem{Frasca:2015yva}
  M.~Frasca,
  Eur.\ Phys.\ J.\ Plus \textbf{132}, no.1, 38 (2017)\\\
  [erratum: Eur.\ Phys.\ J.\ Plus \textbf{132}, no.5, 242 (2017)]

\bibitem{Frasca:2016sky}
  M.~Frasca,
  Eur.\ Phys.\ J.\ C \textbf{77}, no.4, 255 (2017)

\bibitem{Frasca:2017slg}
  M.~Frasca,
  Nucl.\ Part.\ Phys.\ Proc.\ \textbf{294-296}, 124-128 (2018)

\bibitem{Chaichian:2018cyv}
  M.~Chaichian and M.~Frasca,
  Phys.\ Lett.\ B \textbf{781}, 33-39 (2018)

\bibitem{Koberinski:2019tqk}
  A.~Koberinski,
  Synthese \textbf{198}, no.Suppl 16, 3747-3777 (2021)

\bibitem{Brink:2015frg}
  L.~Brink and K.~K.~Phua,
  {\em Proceedings of the Conference on 60 Years of Yang–Mills Gauge Field
    Theories: C.N. Yang's Contributions to Physics}, Nanyang Technological
  University, Singapore, 25 – 28 May 2015. Edited by: L.~Brink (Chalmers
  University of Technology, Sweden) and K.~K.~Phua (NTU, Singapore). June 2016

\bibitem{daSilva:2002dje}
  J.~C.~da Silva, F.~C.~Khanna, A.~Matos Neto and A.~E.~Santana,\\
  Phys.\ Rev.\ A \textbf{66}, 052101 (2002)

\bibitem{Santos:2019xlx}
  A.~F.~Santos and F.~C.~Khanna,
  Int.\ J.\ Mod.\ Phys.\ A \textbf{34}, no.23, 1950128 (2019)

\bibitem{Trotti:2022knd}
  E.~Trotti, S.~Jafarzade and F.~Giacosa,
  Eur.\ Phys.\ J.\ C \textbf{83}, no.5, 390 (2023)

\bibitem{Frasca:2023uaw}
  M.~Frasca and S.~Groote,
  Symmetry \textbf{16}, no.11, 1504 (2024)

\bibitem{BDJ}
  M.~B\"ohn, A.~Denner and H.~Joos,
  ``Gauge Theories of the Strong and Electroweak Interaction,''
  B.G.~Teubner, Stuttgart, 2001

\bibitem{Frasca:2007uz}
  M.~Frasca,
  Phys.\ Lett.\ B \textbf{670}, 73-77 (2008)

\bibitem{Frasca:2009yp}
  M.~Frasca,
  Mod.\ Phys.\ Lett.\ A \textbf{24}, 2425-2432 (2009)

\bibitem{Frasca:2021mhi}
  M.~Frasca, A.~Ghoshal and S.~Groote,\\
  Nucl.\ Part.\ Phys.\ Proc.\ \textbf{318-323}, 138-141 (2022)

\bibitem{Bellac}
  M.~Le Bellac,
  ``Thermal Field Theory,''	
  Cambridge University Press, Cambridge (2008)

\bibitem{Lucini:2012gg}
  B.~Lucini and M.~Panero,
  Phys.\ Rept.\ \textbf{526}, 93-163 (2013)

\bibitem{Borsanyi:2012ve}
  S.~Borsanyi, G.~Endrodi, Z.~Fodor, S.~D.~Katz and K.~K.~Szabo,
  JHEP \textbf{07}, 056 (2012)

\bibitem{Silva:2013maa}
  P.~J.~Silva, O.~Oliveira, P.~Bicudo and N.~Cardoso,\\
  Phys.\ Rev.\ D \textbf{89}, no.7, 074503 (2014)

\bibitem{Peshier:1995ty}
  A.~Peshier, B.~Kampfer, O.~P.~Pavlenko and G.~Soff,
  Phys.\ Rev.\ D \textbf{54}, 2399-2402 (1996)

\bibitem{Castorina:2011ra}
  P.~Castorina, V.~Greco, D.~Jaccarino and D.~Zappala,
  Eur.\ Phys.\ J.\ C \textbf{71}, 1826 (2011)

\bibitem{LISACosmologyWorkingGroup:2022jok}
  P.~Auclair \textit{et al.} [LISA Cosmology Working Group],
  Living Rev.\ Rel.\ \textbf{26}, no.1, 5 (2023)

\bibitem{LISA:2017pwj}
  P.~Amaro-Seoane \textit{et al.} [LISA],
  ``Laser Interferometer Space Antenna,''\\\
  [arXiv:1702.00786 [astro-ph.IM]]

\bibitem{Janssen:2014dka}
  G.~Janssen, G.~Hobbs, M.~McLaughlin, C.~Bassa, A.~T.~Deller, M.~Kramer,
  K.~Lee, C.~Mingarelli, P.~Rosado and S.~Sanidas, \textit{et al.}
  PoS \textbf{AASKA14}, 037 (2015)

\bibitem{Yagi:2011wg}
  K.~Yagi and N.~Seto,
  Phys.\ Rev.\ D \textbf{83}, 044011 (2011)\\\
  [erratum: Phys.\ Rev.\ D \textbf{95}, no.10, 109901 (2017)]

\bibitem{Schwaller:2015tja}
  P.~Schwaller,
  Phys.\ Rev.\ Lett.\ \textbf{115}, no.18, 181101 (2015)

\bibitem{Breitbach:2018ddu}
  M.~Breitbach, J.~Kopp, E.~Madge, T.~Opferkuch and P.~Schwaller,\\
  JCAP \textbf{07}, 007 (2019)

\bibitem{Fairbairn:2019xog}
  M.~Fairbairn, E.~Hardy and A.~Wickens,
  JHEP \textbf{07}, 044 (2019)

\bibitem{Svetitsky:1982gs}
  B.~Svetitsky and L.~G.~Yaffe,
  Nucl.\ Phys.\ B \textbf{210}, 423-447 (1982)

\bibitem{Gross:1984dd}
  D.~J.~Gross, J.~A.~Harvey, E.~J.~Martinec and R.~Rohm,\\
  Phys.\ Rev.\ Lett.\ \textbf{54}, 502-505 (1985)

\bibitem{Dixon:1985jw}
  L.~J.~Dixon, J.~A.~Harvey, C.~Vafa and E.~Witten,
  Nucl.\ Phys.\ B \textbf{261}, 678-686 (1985)

\bibitem{Dixon:1986jc}
  L.~J.~Dixon, J.~A.~Harvey, C.~Vafa and E.~Witten,
  Nucl.\ Phys.\ B \textbf{274}, 285-314 (1986)

\bibitem{Acharya:1998pm}
  B.~S.~Acharya,
  Adv.\ Theor.\ Math.\ Phys.\ \textbf{3}, 227-248 (1999)

\bibitem{Halverson:2015vta}
  J.~Halverson and D.~R.~Morrison,
  JHEP \textbf{04}, 100 (2016)

\bibitem{Asadi:2021pwo}
  P.~Asadi, E.~D.~Kramer, E.~Kuflik, G.~W.~Ridgway, T.~R.~Slatyer
  and J.~Smirnov,
  Phys.\ Rev.\ D \textbf{104}, no.9, 095013 (2021)

\bibitem{Asadi:2021yml}
  P.~Asadi, E.~D.~Kramer, E.~Kuflik, G.~W.~Ridgway, T.~R.~Slatyer
  and J.~Smirnov,
  Phys.\ Rev.\ Lett.\ \textbf{127}, no.21, 211101 (2021)

\bibitem{Bai:2013xga}
  Y.~Bai and P.~Schwaller,
  Phys.\ Rev.\ D \textbf{89}, no.6, 063522 (2014)

\bibitem{Schwaller:2015gea}
  P.~Schwaller, D.~Stolarski and A.~Weiler,
  JHEP \textbf{05}, 059 (2015)

\bibitem{Halverson:2020xpg}
  J.~Halverson, C.~Long, A.~Maiti, B.~Nelson and G.~Salinas,
  JHEP \textbf{05}, 154 (2021)

\bibitem{Bigazzi:2020phm}
  F.~Bigazzi, A.~Caddeo, A.~L.~Cotrone and A.~Paredes,
  JHEP \textbf{12}, 200 (2020)

\bibitem{Bigazzi:2020avc}
  F.~Bigazzi, A.~Caddeo, A.~L.~Cotrone and A.~Paredes,
  JHEP \textbf{04}, 094 (2021)

\bibitem{Huang:2020mso}
  W.~C.~Huang, M.~Reichert, F.~Sannino and Z.~W.~Wang,\\
  Phys.\ Rev.\ D \textbf{104}, no.3, 035005 (2021)

\bibitem{Wang:2020zlf}
  X.~Wang, F.~P.~Huang and X.~Zhang,
  ``Bubble wall velocity beyond leading-log approximation
  in electroweak phase transition,''
  [arXiv:2011.12903 [hep-ph]]

\bibitem{Kang:2021epo}
  Z.~Kang, J.~Zhu and S.~Matsuzaki,
  JHEP \textbf{09}, 060 (2021)

\bibitem{Yamada:2022aax}
  M.~Yamada and K.~Yonekura,
  Phys.\ Lett.\ B \textbf{838}, 137724 (2023)

\bibitem{Yamada:2022imq}
  M.~Yamada and K.~Yonekura,
  Phys.\ Rev.\ D \textbf{106}, no.12, 123515 (2022)

\bibitem{Morgante:2022zvc}
  E.~Morgante, N.~Ramberg and P.~Schwaller,
  Phys.\ Rev.\ D \textbf{107} (2023) no.3, 036010

\bibitem{Bate:1953}
  H.~Bateman and Bateman Manuscript Project,
  ``Higher Transcendental Functions [Volumes I-III],''
  McGraw-Hill Book Company (1953)

\bibitem{WW:1927}
  E.~T.~Whittaker and G.~N.~Watson,
  ``A Course of Modern Analysis. 4th edition,''
  Cambridge University Press (1927)

\bibitem{NIST}
  F.~W.~J.~Olver, A.~B.~Olde Daalhuis, D.~W.~Lozier, B.~I.~Schneider,
  R.~F.~Boisvert, C.~W.~Clark, B.~R.~Miller, B.~V.~Saunders, H.~S.~Cohl, and
  M.~A.~McClain (eds.),
  ``NIST Digital Library of Mathematical Functions,''
  {\tt https://dlmf.nist.gov/}, Release 1.1.6 of 2022-06-30 

\bibitem{TheTables}
  I.~S.~Gradshteyn and I.~M.~Ryzhik,
  ``Table of integrals, series, and products, Seventh edition,''
  translated from the Russian, Translation edited and with a preface by Alan
  Jeffrey and Daniel Zwillinger. Elsevier/Academic Press, Amsterdam (2007)

\bibitem{Hindmarsh:2005ix}
  M.~Hindmarsh and O.~Philipsen,
  Phys.\ Rev.\ D \textbf{71}, 087302 (2005)

\bibitem{Drees:2015exa}
  M.~Drees, F.~Hajkarim and E.~R.~Schmitz,
  JCAP \textbf{06}, 025 (2015)

\bibitem{Hajkarim:2019csy}
  F.~Hajkarim, J.~Schaffner-Bielich, S.~Wystub and M.~M.~Wygas,\\
  Phys.\ Rev.\ D \textbf{99}, no.10, 103527 (2019)

\end{thebibliography}
\end{document}